\documentclass[aps,preprint]{revtex4-1} 
\usepackage{amsmath,amssymb,graphicx,bbm,subcaption,xcolor,braket,latexsym,microtype,bm}
\def\ordNL{\zeta}
\def\Tr{\hbox{Tr}}
\begin{document}
\title{Quantum probes for the characterization of nonlinear media}
\author{Alessandro~Candeloro\,$^{1,2}$, \, Sholeh~Razavian,$^{3,4}$\,  Matteo~Piccolini\,$^{5}$, Berihu~Teklu\,$^{6}$, Stefano~Olivares\,$^{1,2}$ and Matteo~G. A.~Paris,$^{1,2}$}
\address{
$^{1}$ Quantum Technology Lab, Dipartimento di Fisica ``Aldo Pontremoli'', Universit\`a degli Studi di Milano, I-20133 Milano, Italy\\
$^{2}$ INFN, Sezione di Milano, I-20133 Milano, Italy \\
$^{3}$ Max-Planck-Institut fur Quantenoptik, D-85748, Garching bei Munchen, Germany \\
$^{4}$ Department fur Physik, Ludwig-Maximilians-Universit\"at, D-80799 Munchen, Germany \\
$^{5}$ Dipartimento di Ingegneria, Universit\`a di Palermo, I-90128 Palermo, Italy\\
$^{6}$ Department of Applied Mathematics and Sciences and Center for Cyber-Physical Systems (C2PS), Khalifa University, 127788, Abu Dhabi, United Arab Emirates}
\begin{abstract}
Active optical media leading to interaction Hamiltonians of the form $ H = \tilde{\lambda}\, (a + a^{\dagger})^{\ordNL}$ represent a crucial resource for quantum optical technology. In this paper, we address the characterization of those nonlinear media using quantum probes, as opposed to semiclassical ones. In particular, we investigate how squeezed probes may improve {\em individual} and {\em joint} 
estimation of the nonlinear coupling $\tilde{\lambda}$ and of the nonlinearity order $\ordNL$. Upon using tools from quantum estimation, we show that: i) the two parameters are compatible, i.e. the may be jointly estimated without additional quantum noise; ii) the use of squeezed probes improves precision at fixed overall energy of the probe; iii) for low energy probes, squeezed vacuum represent the most convenient choice, whereas for increasing energy an optimal squeezing fraction 
may be determined; iv) using optimized quantum probes, the scaling of the
corresponding precision with energy improves, both for individual and joint estimation 
of the two parameters, compared to semiclassical coherent probes. We conclude that quantum probes represent
a resource to enhance precision in the characterization of nonlinear media, 
and foresee potential applications with current technology.\end{abstract}
\maketitle
\section{Introduction}
Squeezed states and entangled pairs of photons are crucial resources in current implementations of quantum technologies \citep{BROWNE20172}, including 
quantum enhanced sensing, quantum repeaters and the realization of quantum gates in several platforms. The experimental generation of these states exploits the nonlinear response of active materials. In turn, the precise characterization of the nonlinear behaviour of active optical media represents a crucial tool for the development of novel and reliable sensors, aimed at improving protocols for, e.g., non-invasive diagnosis and secure communication.
\par
The quantitative characterization of the nonlinear coupling may be in principle achieved using semiclassical probes, e.g. laser beams in optical systems \citep{Asselberghs2006}, or thermal perturbations in optomechanical ones \citep{Brawley2016,bru11}. 
On the other hand, quantum probes, i.e. probes with nonclassical properties, are naturally  very sensitive to the environment, and can be therefore used to improve precision and make very accurate sensors. As a result of steady progress in material quality and control, cost reduction and the miniaturisation of components, these devices are now ready to be carried over into numerous applications.
\par
From a metrological point of view, the problem of designing a characterization 
scheme for the nonlinearities is twofold. On the one hand, one should find the 
optimal measurement and evaluate the corresponding ultimate bounds to precision: this will serve as a benchmark in the design of any device using nonlinear media. On the other hand, it is necessary to determine the optimal probe signals among those achievable with current technology.
\par
In this paper, we are going to address the above problems for nonlinear interactions corresponding to Hamiltonians of the form $H =\tilde{\lambda}(a+a^{\dagger})^\ordNL$, where $a$ is the annihilation bosonic field operator, $\left[\hat{a}, \hat{a}^\dagger\right]=\mathbb{I}$. In particular, we consider situations where both the 
{\em coupling parameter} $\tilde{\lambda}$ and the {\em order of nonlinearity} 
$\ordNL$ are to be estimated by probing the medium with suitable optical 
signals. These Hamiltonians are encountered rather commonly in quantum optics, 
and provide an effective description of the interaction between 
radiation and matter. In fact, they follow from the quantum interaction 
between a quantized single-mode field and an active medium treated 
parametrically \citep{MW95}.
\par
As a matter of fact, the larger is the nonlinear order, the less effective is 
the nonlinearity. For instance, the non-linear processes naturally occurring in the optical fibers are tiny. On the other hand, they can grow and become relevant as the length of the fiber and, thus, the interaction time, increases. Effects are particularly important in single-mode fibres, in which the small field-mode dimension results in substantially high light intensities despite relatively modest input powers \citep{PhysRevLett.57.691}. In turn, a long-standing goal in optical science has been the implementation of non-linear effects at progressively lower light powers or pulse energies \citep{Andersen_2016}. 
\par
In this paper, our aim is to investigate how the precision of the estimation scales as a function of the average number of photons of the probe, and to assess the performance of different probing signals, with the goal of quantifying the improvement achievable by using nonclassical resources as squeezing. Indeed, there have been several indications in the recent years \citep{qp1,qp2,qp3,qp4,gebbia20} that quantum probes offer advantages in terms of precision and stability compared to their classical counterparts.
In particular, upon using tools from quantum estimation theory \citep{ParisQEQT09,ALBARELLI2020126311}, we are going to determine the optimal measurement to be performed at the output, and to evaluate the corresponding ultimate quantum limit to precision. Additionally, we will investigate the performance of different probe preparations in order to assess whether a nonclassical preparation of the probe may improve precision in some realistic scenarios. 
\par
Our results may find applications in different fields ranging from quantum optics to optomechanics and to more general systems involving phonons \citep{qphon}. Nonetheless, in order to make the presentation more concrete, we will mostly refer to a light beam interacting with optical media. In particular, to illustrate the basic features of our proposal, 
we consider two kinds of probes: customary coherent signals and squeezed ones. 
We let the probe interact with the nonlinear medium, and then we perform a measurement in order to extract information about the parameters we want to estimate. Finally, we evaluate the corresponding quantum Fisher information (QFI) and we determine the optimal probe preparation.
Our findings prove that squeezing is indeed a resource to enhance characterization at the quantum level, especially for fragile samples where a strong constraint on the probe energy is present.
\par
The paper is structured as follows. In Section~\ref{s:MultiPar}, we briefly review the tools of quantum estimation theory. We obtain the ultimate bounds to precision in Section~\ref{s:bounds}, and illustrate our results in Sections~\ref{s:Gprobes} and \ref{s:Gsimprobe}, where we discuss optimal estimation for separate and joint estimation, respectively. Finally, Section~\ref{s:cocnl} closes the paper with some concluding remarks.
\section{Local multiparameter quantum estimation theory}\label{s:MultiPar}
In this section, we introduce the basic tools of local multiparameter quantum estimation theory \citep{Liu_2019, ALBARELLI2020126311}, whose goal is to find the ultimate bounds to  precision in the joint estimation of a finite set of parameters $\left\{ \lambda_n \right\}_{n=1}^{d}$.
\par
The first level of optimization is in the classical setting: in order to maximize the information on the parameters that can be extracted from a collection of experimental data $\bm{X}=(x_1,x_2,...,x_M)$, we need to find a set of optimal estimators, i.e. a set of  maps $\hat{\lambda}_n:\bm{X} \to \mathcal{M}_{\lambda_n}$, where $\mathcal{M}_{\lambda_n}$ is the set of possible values of $\lambda_n$. The usual figure of merit used to assess the goodness of a set of unbiased estimator $\hat{\lambda}_n$ is the  covariance matrix
\begin{equation}
\left[ \bm{V}(\hat{\bm\lambda};\bm\lambda)\right]_{nm}=\int d\bm X\, p(\bm X|\bm \lambda)
\left[ \hat{\lambda}(\bm X)_n-\bar{\lambda}_n \right]
\left[ \hat{\lambda}(\bm X)_m-\bar{\lambda}_m \right],
\end{equation}
where $p(\bm{X}\vert\bm\lambda)$ is the probability distribution of the outcomes when the parameters have values $\bm\lambda$, while $\bar{\bm\lambda}$ is the vector of mean values $\bar{\lambda}_n$ evaluated on $\bm{X}$, i.e. $\bar{\bm\lambda}=\int d\bm X\, p(\bm X|\lambda)\,\hat{\bm\lambda}(\bm X)$. If we introduce the 
Fisher information matrix (FIM)
\begin{equation}
\left[ \bm{\mathcal{F}}^{(c)}_M(\bm\lambda) \right]_{nm}	= \int d\bm X  p(\bm X|\bm\lambda)\left[\partial_{\lambda_n} \log p(\bm X|\bm\lambda)\right]\left[\partial_{\lambda_m} \log p(\bm X|\bm\lambda)\right],
\end{equation}
the ultimate limit of the covariance matrix follows from the request that the matrix $\bm{V}(\lambda)-{\bm{\mathcal{F}}^{(c)}}^{-1}_M(\bm\lambda)$ should be semi-definite positive, that leads to the matrix Cram\'er-Rao inequality
\begin{equation}\label{MCRI}
\bm{V}(\hat{\bm\lambda};\bm\lambda)\geq{\bm{\mathcal{F}}^{(c)}}^{-1}_M(\bm\lambda).
\end{equation}
\par
An important property of the Fisher information is the additivity for 
independent measurements: if the outcomes $x_k$ are independent, then the probability distribution can be factorized as $p(\bm X|\bm\lambda)=\prod_{k=1}^M\,p(x_k|\bm\lambda)$, and thus the FIM becomes $\bm{\mathcal{F}}^{(c)}_M(\bm\lambda)= M \bm{\mathcal{F}}^{(c)}(\bm\lambda)$. Henceforth, we will consider only the scenario where our outcomes are all independent. It is proved that the inequality (\ref{MCRI}) can be always attained in the limit of $M\to\infty$ by a max-likelihood estimator $\hat{\lambda}_{\rm ML}$.
\par
So far we have considered only the classical setting, in which the probability distribution $p(\bm X|\bm\lambda)$ is fixed. On the other hand, the mathematical formalism of quantum mechanics allow us to optimize precision over the full set of possible measurements, thus leading to the ultimate bounds on the attainable precision. In the single parameter scenario, a further optimization among all the possible measurement can be analytically performed in general. This leads to the single-parameter quantum Cram\'er-Rao inequality
\begin{equation}
\label{eq:QFI}
	\mathcal{F}^{(c)}(\lambda) \leq \mathcal{F}^{(q)}(\lambda) = \Tr\left[\rho_\lambda \hat{L}_\lambda^2\right],
\end{equation}
where $\rho_\lambda$ is the density operator of the system, and where we have introduced $\mathcal{F}^{(q)}(\lambda)$,
 the quantum Fisher information (QFI) . The QFI represents the ultimate bound on the precision among the set of all the possible measurements, in general described by a positive-operator valued measure (POVM). Its definition is given in terms of $\hat{L}_\lambda$, the symmetric logarithmic derivative (SLD), which is Hermitian and implicitly defined by the Lyapunov equation \citep{helstrom1976quantum}
\begin{equation}
\label{eq:SLD} 
	2\partial_\lambda\varrho_{\lambda}=\hat{L}_\lambda\,\varrho_{\lambda} + \varrho_{\lambda}\,\hat{L}_\lambda\,. 
\end{equation} 
The SLD is not only essential in the calculation of the QFI, but it is also the key quantity in the determination of the optimal measurement: the projectors of $\hat{L}_\lambda$ correspond to the POVM elements of the optimal measurement.
\par
Once we move to the multiparameter scenario, things change drastically. In principle, we can associate a SLD operator $\hat{L}_{\lambda_n}$ with the corresponding parameter $\lambda_n$, thus we can straightforwardly generalize the QFI in Eq.~\eqref{eq:QFI} to a QFI matrix
\begin{equation}
\label{eq:SLDQFI}
\left[ \bm{\mathcal{F}}^{(q)} (\bm{\lambda}) \right]_{nm}	= \frac{1}{2}\Tr\left[\rho_{\bm\lambda} \{\hat{L}_n,\hat{L}_m\}\right],
\end{equation} 
with $\{\hat A,\hat B\} = \hat A \hat B + \hat B \hat A$. Therefore, any FIM, as well as any covariance matrix $\bm{V}(\bm{\hat{\lambda}},\bm\lambda)$, is lower bounded as
\begin{equation}
\label{eq:SLDQFIbound}
	\bm{V}(\bm{\hat{\lambda}},\bm\lambda) \geq {\bm{\mathcal{F}}^{(c)}}^{-1} (\bm{\lambda})\geq {\bm{\mathcal{F}}^{(q)}}^{-1} (\bm{\lambda})\,.
\end{equation}
If we now introduce the $d\times d$ real, weight matrix $\bm{W}$, we may o
btain the following relation between scalar quantities:
\begin{equation}
\label{eq:SLDQFIboundscalar}
\Tr \left[\bm{W}\bm{V}(\bm{\hat{\lambda}},\bm\lambda)\right] \geq \Tr \left[ \bm{W}{\bm{\mathcal{F}}^{(q)}}^{-1} (\bm{\lambda}) \right] =
C_S(\bm{W},\bm\lambda)\,,
\end{equation}
which takes the name of SLD-QFI bound. The question naturally arises as to whether or not these boundaries are achievable in practice. Clearly, if the matrix bound is attained, also the scalar bound will be, and for this reason we consider the attainability of scalar and matrix bounds as a unique problem \citep{Jordan2019sat}. 
\par
 The goal of multiparameter estimation is to estimate each parameter simultaneously by a single measurement. Therefore, if the SLDs $\{\hat{L}_{\lambda_n}\}_{n=1}^{d}$ do not commute, then the strategy for the optimal estimation for each single parameter $\lambda_n$ can not be performed simultaneously, and the bounds \eqref{eq:SLDQFIbound}-\eqref{eq:SLDQFIboundscalar} are not attainable. However, the achievability of such bounds is subject to a weaker condition which involves the Uhlmann matrix \citep{Carollo2018}
\begin{equation}
\big[ \bm{\mathcal{U}}(\bm\lambda) \big]_{nm} = -\frac{i}{2}\Tr\left\{\rho_{\bm\lambda} \left[\hat{L}_n,\hat{L}_m \right]\right\}.
\end{equation}
The weak compatibility condition states that if $\bm{\mathcal{U}}(\bm\lambda) = \bm{0}$, then the SLD-QFI bound can be attained by an asymptotic statistical model, i.e. by a collective measurement on an asymptotically large number of copy of the state $\rho_{\bm\lambda}$ \citep{Hayashi2019attaining}.
\par
The above expressions can be further simplified in the case of a family of pure states $\rho_{\bm\lambda}=\vert\psi_{\bm\lambda}\rangle\langle\psi_{\bm\lambda}\vert$, in which the SLD can be simply evaluated. Since  $\varrho_{\bm\lambda}^2=\varrho_{\bm\lambda}$, it follows from a direct calculation that $\partial_{\lambda_n}\varrho_{\bm\lambda}=(\partial_{\lambda_n}\varrho_{\bm\lambda})\varrho_{\bm\lambda}+\varrho_{\bm\lambda}(\partial_{\lambda_n}\varrho_{\bm\lambda})$. Hence, from Eq.~({\ref{eq:SLD}}) we easily derive the SLD operator for $\lambda_n$, i.e., $\hat{L}_{\lambda_n}=2\partial_{\lambda_n}\varrho_{\bm\lambda}$. The QFI matrix and the Uhlmann matrix simplifies as well, and we eventually obtain
\begin{subequations}
\begin{align}
\label{eq:QFIpure}
\left[\bm{\mathcal{F}}^{(q)}(\bm\lambda)\right]_{nm}& = 4\Re\mbox{e}\big[\langle\partial_{\lambda_n}\psi_{\bm\lambda}\vert\partial_{\lambda_m}\psi_{\bm\lambda}\rangle+ \langle \partial_{\lambda_n}\psi_{\bm\lambda}\vert \psi_{\bm\lambda}\rangle \langle \partial_{\lambda_m}\psi_{\bm\lambda}\vert \psi_{\bm\lambda}\rangle\big]\,, \\[1ex]
\label{eq:Uhlmpure}
\big[ \bm{\mathcal{U}}(\bm\lambda)\big]_{nm} & = 4\Im\mbox{m}\big[\langle\partial_{\lambda_n}\psi_{\bm\lambda}\vert\partial_{\lambda_m}\psi_{\bm\lambda}\rangle\big]\,.
\end{align}
\end{subequations}
A particular case of interest is given by a parameter $\lambda$ encoded in a unitary evolution $\hat{U}_{\lambda}=\exp{(-i\lambda \hat{G})}$, with $\hat{G}$ the corresponding Hermitian generator. In this case, if the initial probe is a pure state $\vert\psi_0\rangle$,  then the evolved state $\vert\psi_\lambda\rangle = \hat{U}_{\lambda} \vert\psi_0\rangle$ will be pure as well. Hence, we eventually find that the QFI given by Eq.~\eqref{eq:QFIpure} can be expressed in terms of the initial probe and the generator $\hat G$ only as
\begin{equation}
\label{eq:qfipureunitary}
\mathcal{F}^{(q)}(\lambda) = \mathcal{F}^{(q)} = 4\left[\langle\psi_0|\hat{G}^2|\psi_0\rangle -\langle\psi_0|\hat{G}|\psi_0\rangle^2\right]\,,
\end{equation}
namely, it is independent of the parameter, and it is proportional to the fluctuation of $\hat{G}$ on the initial probe $\vert\psi_0\rangle$. Depending on the form of $\hat{G}$, we may be able to optimize the QFI also on $\vert\psi_0\rangle$, obtaining a further optimal bound among all the possible initial probe states. In addition, the SLD operator can be explicitly derived as
\begin{equation}
\label{eq:SLDpure}
	\hat{L}_\lambda = 2 \partial_\lambda \rho_\lambda =2 i \, \hat{U}_\lambda \left[\hat{G},\rho_0 \right] \hat{U}_\lambda^\dagger,
\end{equation}
with $\rho_\lambda  = \vert \psi_\lambda \rangle\langle\psi_\lambda\vert $, from which we can obtain the optimal POVM.
\par
To conclude this brief summary of multiparameter estimation, we consider also how the QFI matrix $\bm{\mathcal{F}}^{(q)}$ is affected by a transformation applied to the parameters. Let us consider a new set of parameters as a function of the formers, namely $\bm\mu = \bm{f}(\bm\lambda)$. Then the new QFI matrix $\bm{\mathcal{F}}^{(q)}(\bm\mu)$ can be expressed in terms of the QFI matrix $\bm{\mathcal{F}}^{(q)}(\bm\lambda)$ as
\begin{equation}
\label{eq:qfichange}
	\bm{\mathcal{F}}^{(q)}(\bm\mu) = \bm{\mathcal{B}} \bm{\mathcal{F}}^{(q)}(\bm\lambda) \bm{\mathcal{B}}^T
\end{equation}
where the matrix $\bm{\mathcal{B}}$ is defined as $[\bm{\mathcal{B}}]_{\mu\nu} = \partial\lambda_\nu/\partial{\mu}_\mu$, where $\bm\lambda = \bm{f}^{-1}(\bm{\mu})$.
\section{QFI matrix for optical non-linearities}\label{s:bounds}
By using the tools of quantum estimation theory, we now find the ultimate 
bounds to precision of estimation of the coupling parameter $\tilde{\lambda}$ and the order $\ordNL$ of a non-linear interaction described by the Hamiltonian
\begin{equation}
\label{eq:Hint}
\hat{H}=\tilde{\lambda}\,\hat{G}_\ordNL\,,
\end{equation}
where  the generator $\hat{G}_\ordNL$ is given by 
\begin{equation}
\label{eq:generator}
	\hat{G}_\ordNL=(\hat{a}+\hat{a}^{\dagger})^\ordNL.
\end{equation}
Accordingly, the time evolution of a pure probe state $\vert\psi_0\rangle$ under the Hamiltonian \eqref{eq:Hint} reads:
\begin{equation}
|\psi_{\bm\lambda}\rangle\equiv|\psi_{\bm\lambda}(t)\rangle=e^{-i\hat{H}\, t}|\psi_0\rangle=e^{-i\lambda\, \hat{G}_\ordNL}|\psi_0\rangle.
\end{equation}
where $\lambda=\tilde{\lambda} t$. Since, by using Eq.~\eqref{eq:qfichange}, we can write
\begin{equation}
	\bm{\tilde{\mathcal{F}}}^{(q)}(\tilde{\lambda},\ordNL)= \bm{\mathcal{B}}\bm{\mathcal{F}}^{(q)}(\lambda,\ordNL)\bm{\mathcal{B}}^T,
\end{equation}
where the matrix elements of $\bm{\mathcal{B}}$ are all null but $[\bm{\mathcal{B}}]_{11}=t$, we can focus only on the joint estimation of $\lambda$ and $\ordNL$, being this totally equivalent to the joint estimation of $\tilde{\lambda}$ and $\ordNL$.
\par
We notice that for the individual estimation of $\lambda$, the element of the QFI matrix is given by Eq.~\eqref{eq:qfipureunitary}, and, from the Hamiltonian \eqref{eq:Hint}, the QFI can be written as
\begin{equation}
\label{QFI}
\mathcal{F}^{(q)}_{\lambda\lambda} = 4 \left[\langle\psi_0| \hat{G}_{2\ordNL}|\psi_0\rangle-\langle\psi_0|\hat{G}_\ordNL |\psi_0\rangle^2\right].
\end{equation}
Analogously, for the estimation of the order of nonlinearity $\ordNL$ only, we have:
\begin{equation}
	\vert \partial_\ordNL \psi_{\bm\lambda}\rangle = -i\lambda \ordNL \hat{G}_{\ordNL-1} \vert \psi_{\bm\lambda} \rangle\,,
\end{equation}
and the corresponding QFI matrix element reads
\begin{equation}
\label{QFI:NL}
	\mathcal{F}^{(q)}_{\ordNL \ordNL} = 4~(\lambda \ordNL)^2\left[\langle\psi_0| \hat{G}_{2(\ordNL-1)}|\psi_0\rangle-\langle\psi_0|\hat{G}_{\ordNL-1}|\psi_0\rangle^2\right].
\end{equation}
By using the expression for $\vert \partial_\ordNL \psi_{\bm\lambda}\rangle$, it is straightforward to evaluate also the off-diagonal elements, obtaining
\begin{equation}\label{mix33}
	\mathcal{F}^{(q)}_{\ordNL\lambda} = 4~\lambda \ordNL \left[\langle \psi_0 \vert \hat{G}_{2\ordNL-1} \vert \psi_0\rangle - \langle \psi_0\vert \hat{G}_\ordNL\vert \psi_0\rangle\langle\psi_0\vert \hat{G}_{\ordNL-1}\vert \psi_0\rangle \right]\,.
\end{equation}
According to the above expressions, the bound to precision for the individual estimation of $\ordNL$ may be derived from that for the estimation of $\lambda$, apart from a rescaling. Together with Eq. (\ref{mix33}) this confirms that all the QFI matrix elements depend on combinations of the expectation value $\langle\psi_0\vert \hat{G}_k\vert\psi_0\rangle$ for different values of $k$, therefore this quantity will be studied in great detail in the following sections.
\par
Regarding the attainability of the QFI-SLD bound \eqref{eq:SLDQFIbound}, this depends on the value of the Uhlmann matrix \eqref{eq:Uhlmpure}. For the statistical model under study, a straightforward calculation shows that the Uhlmann matrix vanishes. Since we are dealing with pure states, we conclude that the model is quasi classical, i.e. joint estimation is possible without additional noise of purely quantum origin and the optimal measurement is given by the projectors of \eqref{eq:SLDpure} for the generator \eqref{eq:generator}.
\section{Optimal probes for individual estimation} \label{s:Gprobes}
After having studied the estimation problem from the point of view of the measurement process, i.e. the QFI matrix corresponding to the optimal measurement, we address now the problem of finding the optimal probe, i.e. the optimal input state to achieve the ultimate bound in the precision of the estimation. In this section, we separately optimize the probe for the individual estimation of $\lambda$ and $\zeta$, i.e. we find the initial states that maximize respectively $\mathcal{F}^{(q)}_{\lambda\lambda}$ and $\mathcal{F}^{(q)}_{\zeta\zeta}$. These optimal probes may not be the same, meaning that different preparations are necessary in order to optimally estimate $\lambda$ or  $\zeta$. The joint estimation of both parameters will be discussed in the next Section.
\par
In our analysis, we focus on the relevant class of Gaussian probes, namely, states that exhibit a Gaussian Wigner function \citep{Olivares2012GS,serafini17book}. In particular, we consider the performance of the so-called \emph{displaced coherent states}, that can be easily generated and manipulated by current quantum optics technology \citep{cialdiPRL20}. Coherent states are usually considered to be the closest quantum states to classical ones. They are eigenstates of the annihilation operator, $\hat{a}|\alpha\rangle=\alpha|\alpha\rangle$, where $\alpha\in\mathbb{C}$, and can be written as
\begin{align}
\vert \alpha\rangle=\hat{D}(\alpha) |0\rangle = e^{-|\alpha|^2/2}\sum_{n}\frac{\alpha^n}{\sqrt{n!}}|n\rangle\,, \label{alphan2}
\end{align}
where $\hat{D}(\alpha)=e^{\alpha a^{\dagger}-\alpha^*a}$ is the displacement operator, $|0\rangle $ the vacuum state and $\{| n \rangle\}_{n\in \mathbbm{N}}$ is the Fock basis.
A displaced squeezed state is defined as follows \citep{Olivares2012GS}
\begin{align}
|\alpha,\xi\rangle = \hat{D}(\alpha)\,\hat{S}(\xi)|0\rangle
\end{align}
where $\hat{S}(\xi)=\exp\left\{\frac{1}{2}\left[\xi(\hat{a}^\dagger)^2 - \xi^* \hat{a}^2\right]\right\}$ is the single-mode squeezing operator and $\xi\in\mathbb{C}$ is the complex squeezing parameter. If $\alpha = 0$, we obtain the so-called \emph{squeezed vacuum} state, whereas for $\xi=0$ we have a coherent state. Given the state $|\alpha,\xi\rangle$, it is convenient 
to introduce the total number of photons $N$ and the {\em squeezing fraction} $\gamma$, namely:
\begin{align}
	N = \langle\alpha,\xi|\hat{N}|\alpha,\xi\rangle= N_{\rm ch}+N_{\rm sq} \quad \mbox{and}\quad
	\gamma =\frac{N_{\rm sq}}{N_{\rm ch}+N_{\rm sq}},
\end{align}
where we set $\xi = r\, e^{i\theta}$, $\hat{N} = \hat{a}^\dagger\hat{a}$ is the number operator and we defined the number of {\em squeezing} photons  $N_{\rm sq} = \sinh^2 r = \gamma N$, whereas the number of {\em coherent} photons  is $N_{\rm ch} = |\alpha|^2 = (1-\gamma)N$. If $\gamma = 0$, we have a coherent state $|\alpha \rangle$, whereas for $\gamma = 1$ we obtain the squeezed vacuum  $|0,\xi\rangle$.
Our ultimate goal is thus determining the optimal parameters $\alpha$ and $\xi$, which realize the maximum of the QFI 
at fixed $N$ and, eventually, to determine the optimal state to probe the non-linear medium in order to estimate the two non-linearity parameters.
\begin{figure}[h!]
 \centering
        \includegraphics[width=0.9\textwidth]{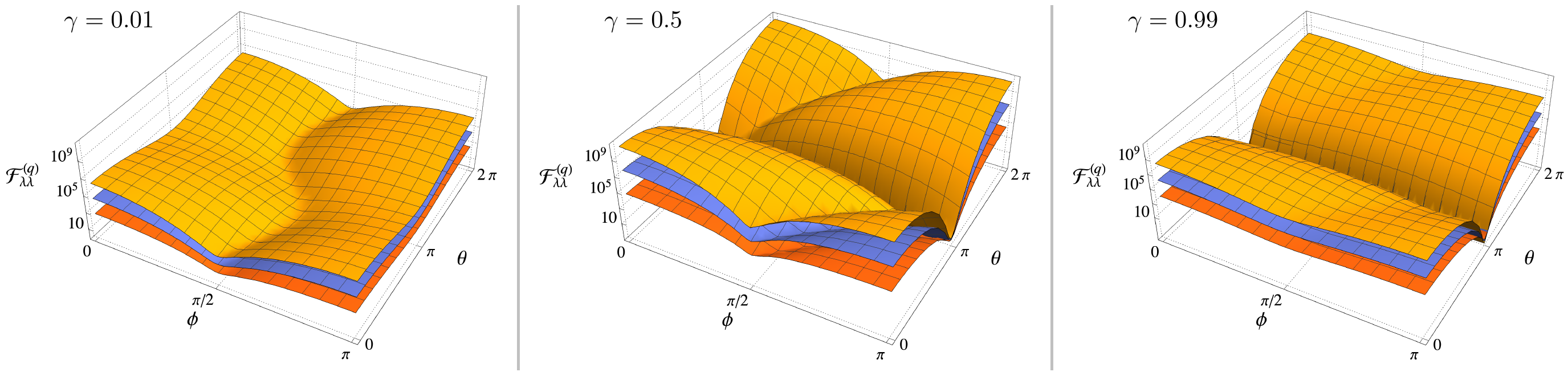}
        \includegraphics[width=0.9\textwidth]{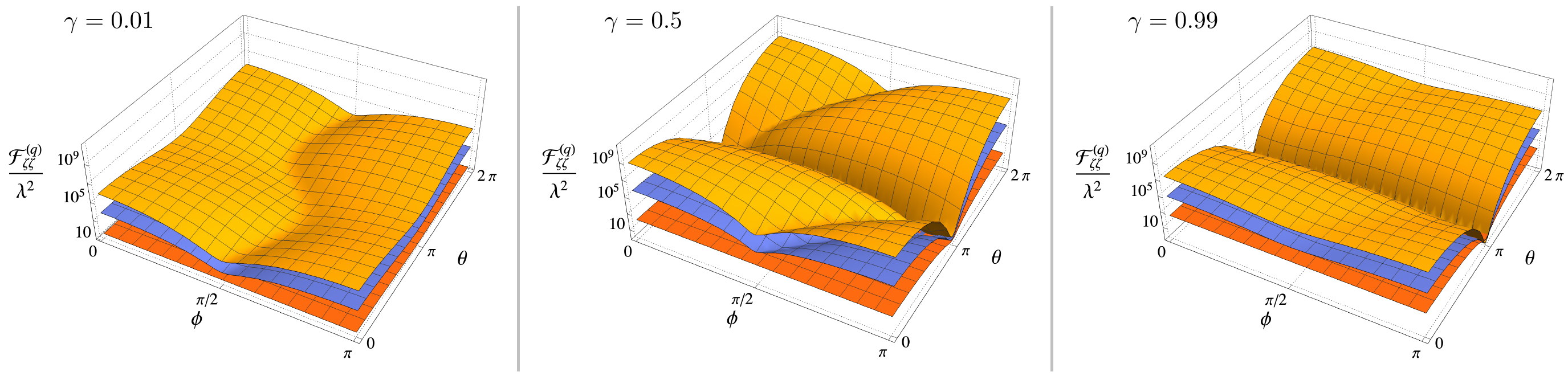}
\caption{\small First line: 
The QFI $\mathcal{F}^{(q)}_{\lambda\lambda}$ of Eq.~(\ref{QFI}) as a function of the squeezing phase $\theta$ and coherent amplitude phase $\phi$ for $N=3$ and for different values of the order of nonlinearity $\ordNL$: from bottom to top $\ordNL=2,3$ and $4$. 
Second line:
The QFI $\mathcal{F}^{(q)}_{\ordNL \ordNL}$ of Eq.~(\ref{QFI:NL}) rescaled by
$\lambda^2$ as a function of the squeezing parameter phase $\theta$ and coherent amplitude phase $\phi$ for $N=3$ and for different values of the order of nonlinearity $\ordNL$: from bottom to top $\ordNL=2,3$ and $4$. 
On both lines, the plots refer to different values of the squeezing ratio: 
(left panels) $\gamma=0.01$ , (middle panels) $\gamma=0.5$ and (right panels) panel: $\gamma=0.99$. 
Notice that the quantity  $\mathcal{F}^{(q)}_{\ordNL \ordNL}/ \lambda^2$ is independent of $\lambda$.}
\label{fig:3d}
 \end{figure}
\par
Following the previous section, given the probe state $|\psi_0\rangle=|\alpha,\xi\rangle$, we have to evaluate the expectation value of $\hat{G}_\ordNL$.  To this aim, we start writing the following identity 
\begin{equation}
	\hat{G}_\ordNL=(\hat{a}+\hat{a}^\dagger)^\ordNL = \ordNL! \sum_{\kappa=0}^{\infty} \delta_{\ordNL \kappa} \frac{1}{\kappa!}(\hat{a}+\hat{a}^\dagger)^\kappa 
\end{equation}
Moreover, we use the following expression for the Kronecker delta
\begin{equation}
	\delta_{\kappa\ordNL} = \frac{1}{2\pi}\int^{\pi}_{-\pi} \!\!\!\! dx\, e^{i(\kappa-\ordNL)x} ,
\end{equation}
which lead us to
\begin{align}
	\hat{G}_\ordNL &= \ordNL!\int^{\pi}_{-\pi} \frac{dx}{2\pi}e^{-i\ordNL x}\sum_{\kappa=0}^{+\infty} \frac{e^{i\kappa x}}{\kappa!}(\hat{a}+\hat{a}^\dagger)^\kappa\\
	& = \ordNL! \int^\pi_{-\pi} \frac{dx}{2\pi} e^{-i\ordNL x} e^{ix(\hat{a}+\hat{a}^\dagger)}.
\end{align}
Now, considering that the creation and annihilation operator satisfy $[\hat{a},\hat{a}^\dagger]=\mathbb{I}$, we can write $e^{ix(\hat{a}+\hat{a}^\dagger)}= \exp\{e^{ix}\hat{a}^\dagger\}\exp\{e^{ix}\hat{a}\}\exp\{e^{2ix}/2\}$, and consequently we obtain
\begin{align}
 	\hat{G}_\ordNL & = \ordNL! \int^\pi_{-\pi}\frac{dx}{2\pi} e^{-i\ordNL x} \sum_{s=0}^{+\infty}\frac{(e^{ix}\hat{a}^\dagger)^s}{s!}\sum_{t=0}^{+\infty}\frac{(e^{ix}\hat{a})^t}{t!}\sum_{m=0}^{+\infty}\frac{(e^{2ix})^m}{2^m m!} = \\
 	& = \ordNL! \sum_{s,t,m=0}^{+\infty} \frac{(\hat{a}^\dagger)^s(\hat{a})^t}{s!t!m!2^m} \int^{\pi}_{-\pi} \frac{dx}{2\pi} e^{ix(s+t+2m-\ordNL)} = \\
 	& = \sum_{s,t,m=0}^{+\infty} \frac{\ordNL!}{s!t!m!2^m} (\hat{a}^\dagger)^s(\hat{a})^t \delta_{\ordNL,s+t+2m}.
\end{align} 
In the last expression, we may perform the sum over $t$ and, noticing that $s$ can be at most $\ordNL-2m$, while $m$ can be at most $\lfloor \ordNL/2 \rfloor$, we finally obtain \citep{wil1,luisell}
\begin{align}
\label{eq:formula1}
\hat{G}_\ordNL = \sum_{m=0}^{\lfloor \ordNL/2\rfloor} \sum_{s=0}^{\ordNL-2m} C(\ordNL,m,s) (\hat{a}^\dagger)^s (\hat{a})^{\ordNL-2m-s}\,,
\end{align} 
where
\begin{align}
C(\ordNL, m, s) =  \frac{\ordNL!}{2^m m!s! (\ordNL-2m-s)!}\,.
\end{align}
More generally, the normal order of $(e^{i\psi}\hat{a}+e^{-i\psi}\hat{a}^\dagger)^\ordNL$ may be obtained. In this case, we redefine the ladder operators as $\hat{b}= e^{i\psi}\hat{a}, \hat{b}^\dagger = e^{-i\psi}\hat{a}^\dagger$, which satisfy the canonical commutation relations $\left[\hat{b},\hat{b}^\dagger\right] = \mathbb{I}$. Then, it results that
\begin{align}
	(e^{i\psi}\hat{a}+e^{-i\psi})^\ordNL & = (\hat{b}+\hat{b}^\dagger)^\ordNL =\\
	& = \sum_{m=0}^{\lfloor \ordNL/2 \rfloor} \sum_{s=0}^{\ordNL-2m} \frac{\ordNL!}{2^m s!m! (\ordNL-s-2m)!}(\hat{b}^\dagger)^s (\hat{b})^{\ordNL-s-2m} = \\
	& = \sum_{m=0}^{\lfloor \ordNL/2 \rfloor} \sum_{s=0}^{\ordNL-2m} \frac{\ordNL!e^{i\psi(\ordNL-2l-2m)}}{2^m s!m! (\ordNL-s-2m)!}(\hat{a}^\dagger)^s (\hat{a})^{\ordNL-s-2m}.
\end{align}
In turn, we have that 
\begin{align}
\langle\alpha,\xi|\,\hat{G}_\ordNL\,|\alpha,\xi\rangle&=
\langle 0|\,\hat{S}^{\dagger}(\xi)\,\hat{D}^{\dagger}(\alpha)\,(\hat{a}+\hat{a}^{\dagger})^\ordNL\, \hat{D}(\alpha)\, \hat{S}(\xi)\,|0\rangle \nonumber\\
&= \langle\beta|\left[(\mu+\nu^*)\hat{a}+(\mu+\nu)\hat{a}^{\dagger}\right]^\ordNL |\beta\rangle 
= \eta^\ordNL \langle\beta\vert\left(\hat{a}e^{i\psi} + \hat{a}^\dagger e^{-i\psi}\right)^\ordNL\vert\beta\rangle = \nonumber \\
&= \eta^\ordNL \sum_{k=0}^{\lfloor \ordNL/2 \rfloor}\,\sum_{s=0}^{\ordNL-2k}\, C(\ordNL,k,s)e^{i\psi(\ordNL-2k-2s)}\,{(\beta^{*})}^s\,\beta^{\ordNL-2k-s},
\label{TUTTO}
\end{align}
where we have introduced $\beta=\mu\alpha+\nu \alpha^*$, $\eta = \vert \mu+\nu\vert$ and $\psi= \textup{Arg}(\mu+\nu^*)$,  with $\mu=\cosh r$ and $\nu=e^{i\theta}\,\sinh r$.  Starting from Eq.~(\ref{TUTTO}) we can evaluate the QFI of Eqs.~(\ref{QFI}) and (\ref{QFI:NL}), which are shown in Figure~\ref{fig:3d}. As one may expect, the behaviour is qualitatively similar, except for the case $\ordNL = 2$ and for $\gamma \to 0$, i.e. for a coherent probe: in this case the QFI associated with the estimation of the order of nonlinearity $\zeta$ does not depend on the parameters of 
the probe state and reads $\mathcal{F}^{(q)}_{\ordNL \ordNL} = 16 \lambda^2$. 
\par
\begin{figure}[h!]
 \centering
        \includegraphics[width=0.8\textwidth]{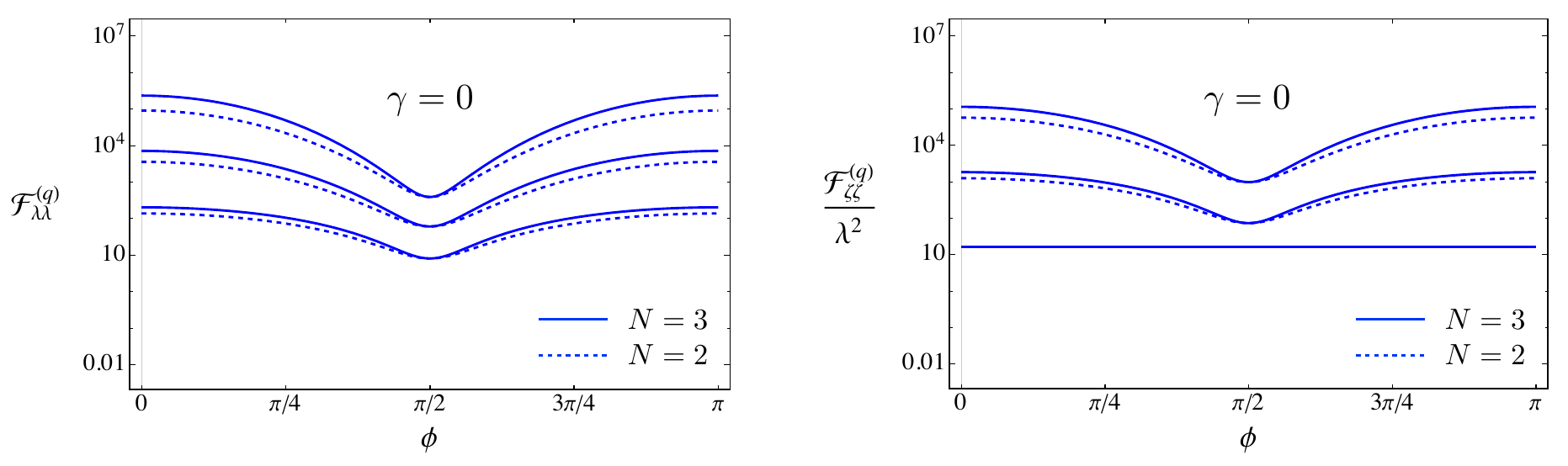}
        \includegraphics[width=0.8\textwidth]{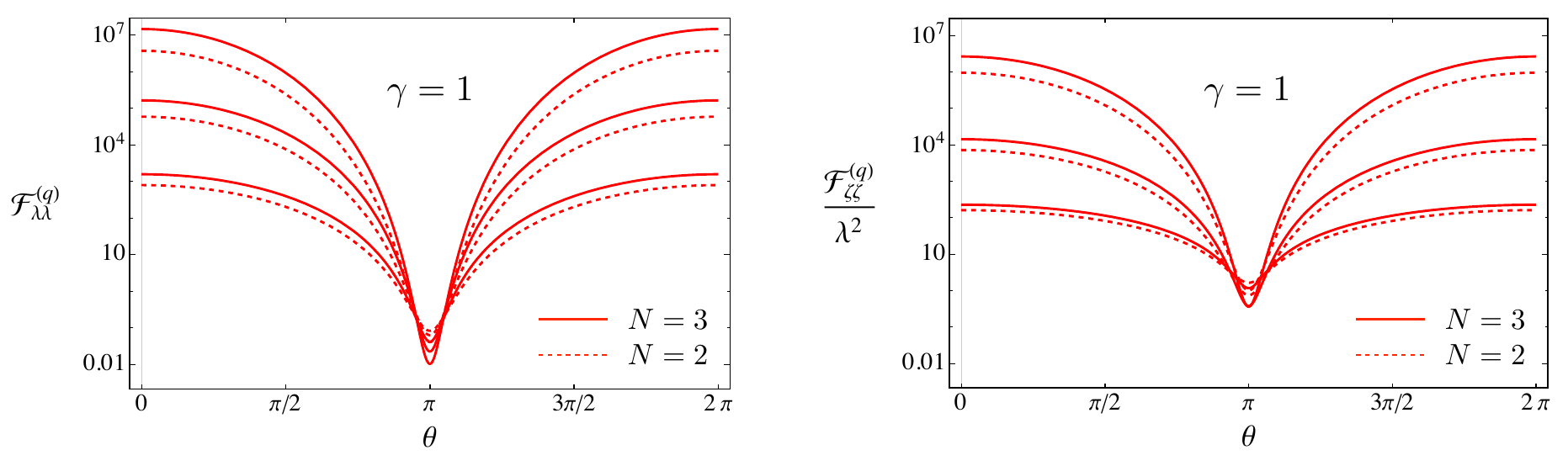}
\caption{\small Upper plots: $\mathcal{F}^{(q)}_{\lambda\lambda}$ and $\mathcal{F}^{(q)}_{\ordNL \ordNL}/ \lambda^2$ for a coherent  probe, i.e. $\gamma=0$, as a function of the coherent state phase $\phi$ for $N = |\alpha|^2 = 2$ (dashed lines) and $N = |\alpha|^2 = 3$ (solid lines) and different values of the order of nonlinearity: form bottom to top $\ordNL=2,3$ and $4$. Note that for $\ordNL = 2$ we have $\mathcal{F}^{(q)}_{\ordNL \ordNL}/ \lambda^2=16$ (lower line the right panel).
Lower plots: $\mathcal{F}^{(q)}_{\lambda\lambda}$ and $\mathcal{F}^{(q)}_{\ordNL \ordNL}/ \lambda^2$ for a squeezed vacuum probe, i.e. $\gamma=1$, as functions of the squeezing  phase $\theta$ for $N = \sinh^2 r = 2$ (dashed lines) and $N = \sinh^2 r = 3$ (solid lines) and different values of the order of nonlinearity: form bottom to top $\ordNL=2,3$ and $4$.}
\label{fig:g01}
 \end{figure}

In Figure~\ref{fig:g01} we show the QFIs for the two extreme cases, i.e 
a coherent probe and a squeezed vacuum one, respectively, as a function of the relevant phases.
From the Figures above,  it is clear that both $\mathcal{F}^{(q)}_{\lambda\lambda}$ and $\mathcal{F}^{(q)}_{\ordNL \ordNL}$ are periodic functions of the phases $\phi$ and $\theta$ of the probe state. Since we are interested in finding the optimal probes, i.e. states maximizing the QFIs, we set $\theta=\phi=0$. Thereafter, we have $\alpha \in \mathbb{R}$, $\beta = \alpha \, e^r$ and  $\eta= e^r$ and Eq.~(\ref{TUTTO}) can be rewritten as
\begin{align}
	\langle\alpha,r |\,\hat{G}_\ordNL\,|\alpha,r \rangle
	& = (\alpha \, e^{2r})^{\ordNL} \sum_{k=0}^{\lfloor \ordNL/2 \rfloor}\,(\alpha \, e^r)^{-2k}
	\sum_{s=0}^{\ordNL-2k}\, C(\ordNL,k,s)\,,
\end{align}
and, being \citep{abramowitz1964handbook}
\begin{equation}
\sum_{s=0}^{\ordNL-2k} C(\ordNL,k,s) = \frac{2^{\ordNL-3k}\ordNL!}{k!(\ordNL-2k)!},
\end{equation}
we eventually obtain:
\begin{equation}
	\langle\alpha,r |\,\hat{G}_\ordNL\,|\alpha,r \rangle = (2\alpha\,e^{2r})^\ordNL \ordNL!\sum_{k=0}^{\lfloor \ordNL/2 \rfloor} \frac{(2\sqrt{2}\,\alpha\,e^{r})^{-2k}}{k!(\ordNL-2k)!}.
\end{equation}
We can now use this last result to evaluate the corresponding QFIs and look for the optimal squeezing fraction $\gamma$ maximizing them.
\par
\begin{figure}[h!]
 \centering
        \includegraphics[width=0.9\textwidth]{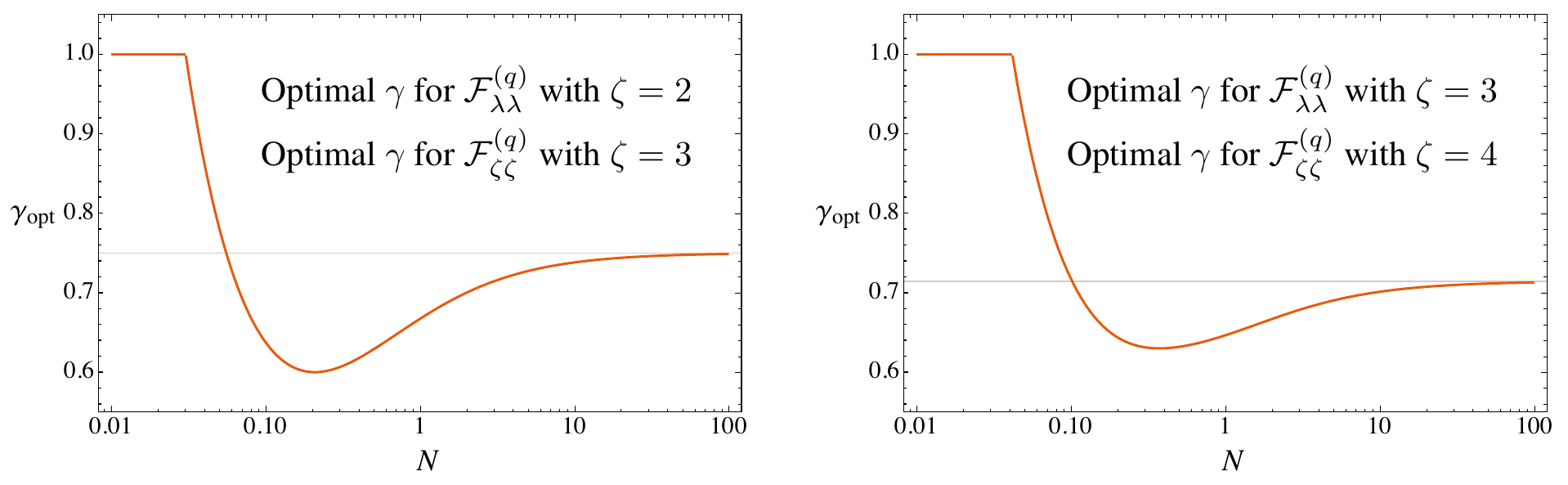}
\caption{\small The optimal squeezing fraction $\gamma_{\rm opt}$ maximizing $\mathcal{F}^{(q)}_{\lambda\lambda}$ and $\mathcal{F}^{(q)}_{\ordNL \ordNL}$ for different values of the nonlinearity order $\ordNL$. The horizontal lines corresponds to the asymptotic value given in Eq.~(\ref{gamma:N:large}). See the text for details.}
\label{f:optimal:gamma}
 \end{figure}

 
At first, we study the low energy $N\ll 1$ regime, where we may write
\begin{equation}
\mathcal{F}^{(q)}_{\lambda\lambda} \simeq 4\, \frac{A(\ordNL)}{2^{\ordNL }} \left(
1 + 2 \ordNL \sqrt{\gamma N} 
\right)
\qquad (N \ll 1)
\end{equation}
and
\begin{equation}
\mathcal{F}^{(q)}_{\ordNL\ordNL} \simeq 4\, \lambda^2 \ordNL^2\,\frac{A(\ordNL-1)}{2^{\ordNL-1}}
\left[
1 + 2 (\ordNL -1) \sqrt{\gamma N} 
\right]
\qquad (N \ll 1)
\end{equation}
where
\begin{equation}
\displaystyle
A(\ordNL) = \left\{
\begin{array}{ll}
{\displaystyle \frac{(2\ordNL)!}{\ordNL !} }& \mbox{if $\ordNL$ odd;}\\[2ex]
{\displaystyle \frac{(2\ordNL)!}{\ordNL !} - \left[ \frac{\ordNL!}{(\ordNL/2)!} \right]^2 }& \mbox{if $\ordNL$ even.}
\end{array}
\right.
\end{equation}
These expansions suggest the existence of a threshold value of $N$, which depends 
on $\ordNL$, below which the QFI reaches the maximum for $\gamma = 1$ (i.e. for a squeezed vacuum probe). Indeed, the maximization at fixed $N$ confirms this intuition. In Figure~\ref{f:optimal:gamma} we show the optimal value of $\gamma$, maximizing the QFIs, as a function of $N$ for two values of $\zeta$.

As we can see from Figure~\ref{f:optimal:gamma}, due to the particular mathematical relations between $\mathcal{F}^{(q)}_{\lambda\lambda}$ and $\mathcal{F}^{(q)}_{\ordNL\ordNL}$, the same optimal squeezing fraction $\gamma_{\rm opt}$ maximizing $\mathcal{F}^{(q)}_{\ordNL\ordNL}$ for a given $\ordNL$ maximizes also $\mathcal{F}^{(q)}_{\lambda\lambda}$ for the order of nonlinearity $\ordNL-1$. We have an exception for $\ordNL=2$: in this peculiar case, to reach the maximum value of $\mathcal{F}^{(q)}_{\ordNL\ordNL}$, one should always choose $\gamma=1$ (squeezed vacuum probe), as we can see by its rather simple analytic expression:
\begin{equation}
\mathcal{F}^{(q)}_{\ordNL\ordNL} = 16 \lambda^2 \left[
1 + 2 \gamma N + 2 \sqrt{\gamma N (1 + \gamma N)}
\right] \qquad (\ordNL = 2).
\end{equation}
Apart from this exception, we observe a threshold value $N_{\rm th}$ for $\gamma_{\rm opt}<1$, i.e the squeezed vacuum is no longer the optimal probes. The values of $N_{\rm th}$ depends on the order of the non-linearity: for the estimation of $\lambda$ and for even values $\zeta$ or for the estimation of $\zeta$ and for odd values of $\zeta$ (left panel of Figure~\ref{f:optimal:gamma}) it is equal to $N_{\rm th} = (3\sqrt{2}-4)/8\simeq 0.03$, while for the other cases (right panel of Figure~\ref{f:optimal:gamma}) the $N_{\rm th}$ approaches $(3\sqrt{2}-4)/8$ for $\zeta \geq 5$. 
\begin{figure}[h!]
 \centering
        \includegraphics[width=0.9\textwidth]{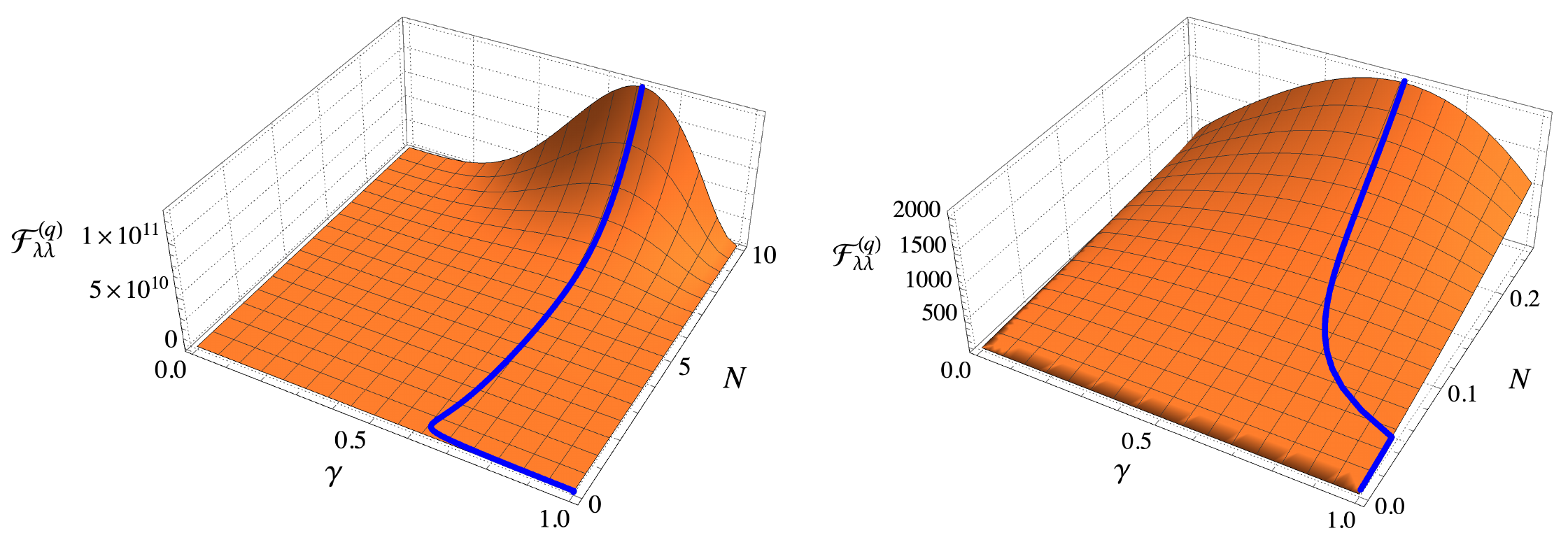}
\caption{\small Plot of $\mathcal{F}^{(q)}_{\lambda\lambda}$ as a function of $\gamma$ and $N$ for $\zeta = 3$. The right panel is a magnification of the left one to highlight the behaviour of the QFI in the regime $N\ll 1$. The blue line refers to the maximum of the QFI (see also the right panel of Figure~\ref{f:optimal:gamma}). Analogous results can be obtained for $\mathcal{F}^{(q)}_{\lambda\lambda}$ and other values of $\ordNL$. See the text for details.}
\label{f:3d:opt}
 \end{figure}
\par
In the large energy regime,  the QFIs are found to grow as  
\begin{align}
\mathcal{F}^{(q)}_{\lambda\lambda} & \simeq B_{\gamma}(\ordNL)\,N^{3\ordNL-2}
\qquad (N \gg 1)
\end{align}
and
\begin{align}
\mathcal{F}^{(q)}_{\ordNL\ordNL} & \simeq \lambda^2 \ordNL^2 B_{\gamma}(\ordNL-1)\,N^{3(\ordNL-1)-2}
\qquad (N \gg 1)
\end{align}
respectively, with
\begin{equation}
B_{\gamma}(\ordNL) =  4^{3\ordNL-1} \ordNL^2 (1-\gamma)^{\ordNL-1}\gamma^{2\ordNL-1}\,.
\end{equation}
Using the results in the large energy regime $N \gg 1$ it is easy to find that the optimal squeezing fraction maximizing $\mathcal{F}^{(q)}_{\lambda\lambda}$ is given by (the optimal squeezing fraction maximizing $\mathcal{F}^{(q)}_{\ordNL\ordNL}$ can be obtained replacing $\ordNL$ with $\ordNL-1$, as it is clear from the previous equations):
\begin{equation}
\label{gamma:N:large}
\gamma_{\rm opt}( N \gg 1)= \frac{2\ordNL-1}{3\ordNL-1}\,,
\end{equation}
and, therefore, $\gamma_{\rm opt}\to 2/3$ as $\ordNL$ increases, as one can also see from Figure~\ref{f:optimal:gamma}.
\par 
We summarize results in Figure~\ref{f:3d:opt}, where we show the QFI as a function of $\gamma$ 
and $N$ for a given value of the order of nonlinearity $\ordNL$. The blue lines denote the 
maxima of the QFI, which are of course obtained for the value of 
hte optimal squeezing ratio $\gamma_{\rm opt}$ displayed in the 
right plot of Figure~\ref{f:optimal:gamma}.
\section{Optimal probes for joint estimation}\label{s:Gsimprobe}
In the previous Section we have individuated the optimal probes for the \emph{individual} estimation of $\lambda$ and $\zeta$, and we have seen that they do not match, i.e.  given a nonlinear media, the optimal probe for the estimation of $\lambda$ may not be optimal for $\zeta$. 
\par
\begin{figure}[h!]
 \centering
        \includegraphics[width=0.9\textwidth]{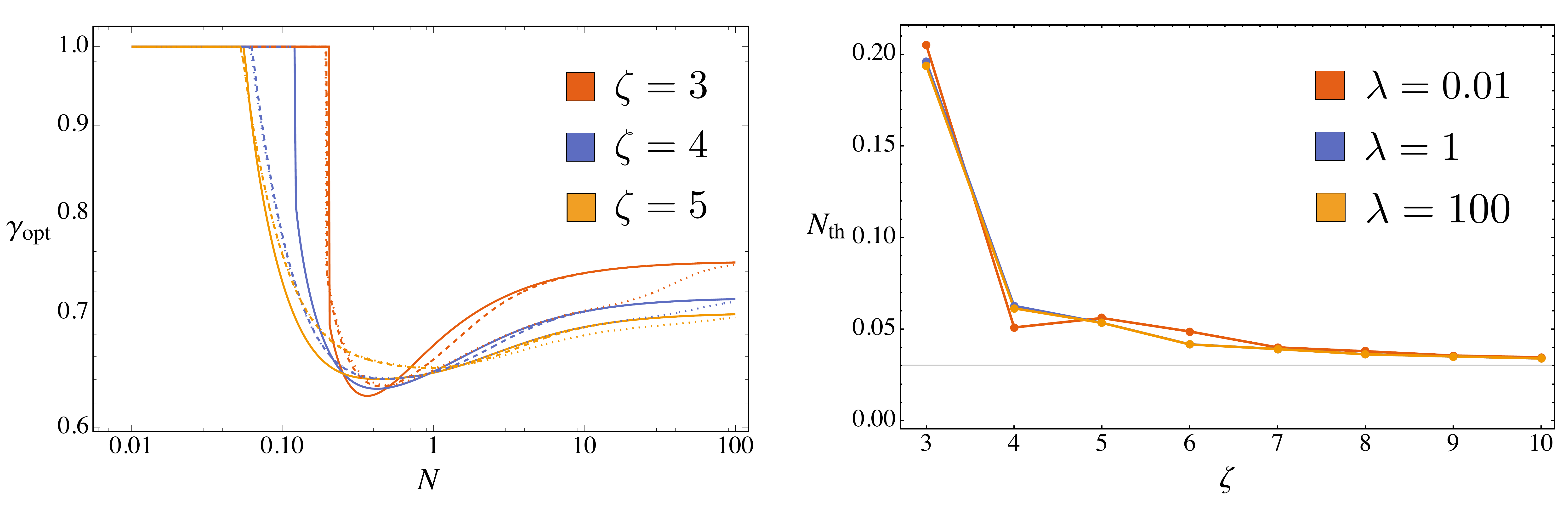}
\caption{\small Left panel: optimal value of the fraction of squeezing $\gamma$ for the scalar bound $C^{-1}_S(\bm{I},\{\lambda,\zeta\})$ as a function of $N$ and for $\lambda =0.01$ (solid lines), $\lambda=1$ (dashed lines) and $\lambda=100$ (dotted lines). Right panel: threshold value $N_{\rm th}$ we observe in the left panel. If $N<N_{\rm th}$ the squeezed vacuum is optimal, otherwise the optimal probe has $\gamma_{\rm opt}<1$.}
\label{f:gamma:opt:nth}
 \end{figure}

In this Section we address the \emph{joint} estimation of both $\lambda$ and $\zeta$ and we find the optimal probe for the multiparameter scenario. In this case, the figure of merit to be maximised is neither the $\mathcal{F}_{\lambda\lambda}$ or the $\mathcal{F}_{\zeta\zeta}$, but the inverse of the scalar bound given in Eq.~\eqref{eq:SLDQFIboundscalar}. For the estimation of two parameters, this can be explicitly evaluated. If we consider the weight matrix to be  $\bm{W}=\mathbb{I}$, i.e. we assume that the estimation of $\lambda$ has the same importance of the estimation of $\zeta$, we eventually obtain
\begin{equation}
	C^{-1}_S(\bm{I},\{\lambda,\zeta\}) = \frac{\mathcal{F}^{(q)}_{\lambda\lambda}\mathcal{F}^{(q)}_{\zeta\zeta}-{\mathcal{F}^{(q)}_{\lambda\zeta}}^2}{\mathcal{F}^{(q)}_{\lambda\lambda}+\mathcal{F}^{(q)}_{\zeta\zeta}}.
\end{equation}
In addition, due to the periodicity of the matrix elements of the QFI matrix, we still focus on the case $\theta=\phi=0$. In this way, we can optimize the inverse of the scalar bound $C^{-1}_S(\bm{I},\{\lambda,\zeta\})$ in a similar way as we did in the previous Section for the individual QFIs. However, here the expression of the scalar bound is more involved, and we have to address the problem numerically. Results are reported in Figure~\ref{f:gamma:opt:nth}. From the left panel, we may see that squeezed vacuum is optimal for $N<N_{\rm th}$, while in the limit of large $N$ the optimal fraction of squeezing $\gamma_{\rm opt}$ depends only on the order of non linearity. Looking at 
the right panel, we see that threshold value $N_{\rm th}$ depends both on $\zeta$ and $\lambda$, even though there are no significant difference for the different values 
of $\lambda$ we have considered. As for the individual estimation, the $N_{\rm th}$ approaches an asymptotic value as the order of non-linearity increases. The value is slightly larger than the one found in the previous section.
\par
\begin{figure}[h!]
 \centering
        \includegraphics[width=0.9\textwidth]{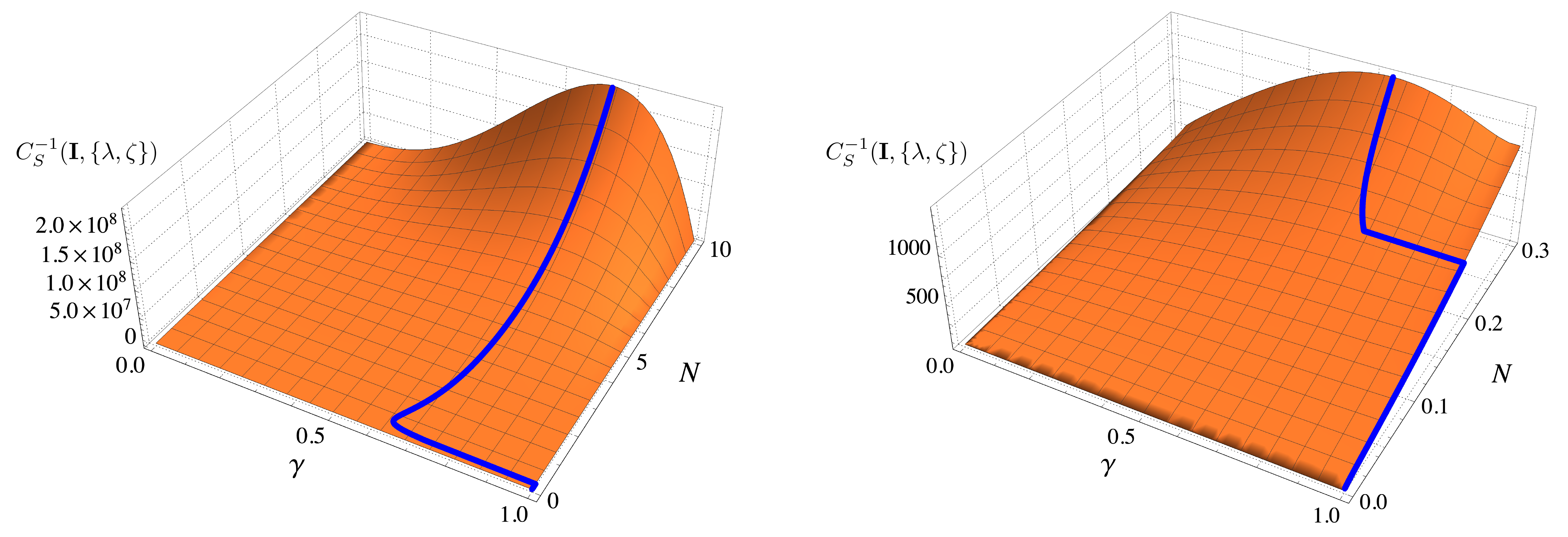}
\caption{\small Plot of $C_S^{-1}(\bm{I},\{\lambda,\zeta\})$ as a function of $\gamma$ and $N$ for $\zeta = 3$. The right panel is a magnification of the left one to highlight the behaviour of the QFI in the regime $N\ll 1$. The blue line refers to the maximum of the QFI at fixed $N$.}
\label{f:3d:opt:CSinv}
 \end{figure}

In Figure~\ref{f:3d:opt:CSinv} we plot the quantity $C^{-1}_S(\bm{I},\{\lambda,\zeta\})$ as a function of $\gamma$ and $N$ and for $\zeta=3$. We have highlighted the optimal value of the scalar bound with a blue lines. Comparing this Figure with the corresponding one for separate estimation (see Figure ~\ref{f:3d:opt}), we see that the qualitative behaviour is the same, while we notice that the $N_{\rm th}$ is slightly larger, as we already outlined in previous considerations. This behaviour can be understood by the fact that we have to find a trade-off between the optimality for $\lambda$ and $\zeta$. 
\section{Conclusions}\label{s:cocnl}
In this paper, we have addressed the use of squeezed states to improve precision in the characterization of nonlinear media. This is inherently a multiparameter estimation problem since it involves both the nonlinear coupling and the order of nonlinearity. 
Using tools from quantum estimation theory we have firstly proved that the two parameters are compatible, i.e. they may be jointly estimated without introducing any noise of quantum origin. In turn, this opens the possibility of exploiting squeezing as a resource to overcome the limitation of coherent probes. 
\par
We have found that using squeezed probes improves the estimation precision in any working regime, i.e. either for fragile media where one is led to use low energy probes, or when this constraint is not present, and one is free to choose probes with high energy. In the first case, squeezed vacuum represents a universally optimal probe \citep{par95,gaiba}, where, for higher energy, squeezing should be tuned and depends itself on the value of the nonlinearity. This results hold both for the separate estimation of the two parameters, as well as for their joint estimation.
In all regimes, using squeezing improves the scaling of the precision with the energy of the probe.
\par
We conclude that quantum probes exploiting squeezing are indeed a resource for the characterization of nonlinear media. {Actually, this involves a more complex probe preparation compared to the semiclassical case. However, in view of the current development in quantum optics, we foresee potential applications with current technology}.
\section*{Conflict of Interest Statement}
The authors declare that the research was conducted in the absence of any commercial or financial relationships that could be construed as a potential conflict of interest.
\section*{Funding}
This work has been supported by Khalifa University through project 
no. 8474000358 (FSU-2021-018) and by MAECI through the project ``ENYGMA'' no. PGR06314.

\section*{Acknowledgments}
MGAP is member of INdAM-GNFM.

\end{document}